\documentclass{amsart}  
\usepackage[english]{babel}
\usepackage[utf8]{inputenc}
\usepackage{algorithmic}
\usepackage{algorithm}
\usepackage{amsfonts}
\usepackage{amsmath}
\usepackage{amssymb}
\usepackage{amsthm}
\usepackage{enumerate}
\usepackage{epsfig}
\usepackage{graphicx}
\usepackage{amsaddr}
\usepackage{setspace}
\usepackage{subfigure}
\usepackage{url}

\newcommand{\wv}{\textbf{w}}
\newcommand{\xv}{\textbf{x}}

\newcommand{\st}{\mbox{s.t.}}

\newcommand{\Am}{\textbf{A}}

\newcommand{\Cm}{\textbf{C}}
\newcommand{\Dm}{\textbf{D}}

\newcommand{\Rm}{\textbf{R}}

\newcommand{\Xm}{\textbf{X}}

\newcommand{\Zm}{\textbf{Z}}

\begin{document}

\title{On Recent Advances in Supervised Ranking \\for Metabolite Profiling}

\author{Charanpal Dhanjal \and St\'ephan Cl\'{e}men\c{c}on}
\address[Charanpal Dhanjal]{Institut Mines-T\'{e}l\'{e}com; T\'{e}l\'{e}com ParisTech, CNRS LTCI, 46 rue Barrault, 75634 Paris Cedex 13, France}
\email[Charanpal Dhanjal]{\{charanpal.dhanjal, stephan.clemencon\}@telecom-paristech.fr}

\date{\today}

\setlength{\tabcolsep}{0.5pt}

\begin{abstract}
\noindent This paper focuses on data arising from the field of \textit{metabolomics}, a rapidly developing area concerned by the analysis of the chemical fingerprints (\textit{i.e.} the metabolite profile). The metabolite profile is left by specific chemical processes occurring in biological cells, tissues or organs. It is the main purpose of this article to develop and implement scoring techniques so as to rank all possible \textit{metabolic profiles} by increasing order of magnitude of the conditional probability that a given metabolite is present at high levels in a certain biological fluid.  After a detailed description of the (functional) data from which decision rules must be learnt, several approaches to this predictive problem, based on recent advances in $K$-partite ranking are described at length. Their performance on several real datasets are next thoroughly investigated.
\end{abstract}

\maketitle


\section{Introduction}

Thanks to high resolution \textit{Mass Spectrometry} ({\sc MS}) and \textit{Nuclear Magnetic Resonance} (NMR), it is now possible to get very detailed metabolic information about the chemical processes which might occur in a biological cell, tissue, organ or organism. \textit{Metabolomics} is precisely the study of these biological processes through the quantitative measurements obtained by means of one of these analytical platforms. It  belongs to the most recent -omics techniques used in order to \textit{phenotype} micro-organisms or multicellular organisms like plants or animals, \textit{i.e.} to characterise the metabolic response of such living systems to given pathophysiological stimuli or genetic modification, see \cite{nicholson1999metabonomics, fernie2004metabolite, roux2010applications} for instance.

Whereas {\sc MS} measures the relative abundance in ionised analytes, directly or after a chromatographic step, which prevents some signal extinction problems and warrants a better quantification of the few ionised species simultaneously detected at a given retention time, {\sc NMR} is less sensitive but more rapid to implement and needs a very simple sample preparation (the term ``metabonomics'' is sometimes used to distinguish analyses based on NMR measurements). NMR detection of analytes corresponds to the detection of the spin of non-exchangeable protons, or carbon-13, present on analytes. Due to the presence of chemical neighbouring of specific chemical functions, assignments of these spins is specific for a given analyte. Nevertheless, in biological matrices, due to the coalescence of some signals coming from numerous analytes in certain spectral regions, the ultimate identification can be difficult to obtain in one dimensional NMR. For any individual submitted to this phenotyping procedure, this information is summarised inside a spectroscopic fingerprint, which contains hundreds to thousands of analytes. They are a priori considered as independent variables. Mining such a large information requires sophisticated statistical tools \cite{trygg2007chemometrics, eliasson2011data, ebbels2011processing}. This represents a highly challenging domain in statistics research, most tools being still at a conceptual or a developing stage. 

Contrary to MS-based fingerprints, NMR ones can be obtained easily on very large populations with a high reproducibility and therefore are well adapted to an epidemiological survey of the general population or of some targeted groups, which display some pathological risks \cite{holmes2008metabolic}. Indeed, this is the case of high level or elite sportsmen who are submitted in France since beginning of 2000s to the longitudinal medical follow-up to detect endocrine anomalies which could be indicative of forbidden doping practices. In bike athletes, a biobank containing sera of sportsmen submitted to this longitudinal follow-up has been progressively built. In this biobank are present some sera in which extreme (lower or higher than normal) values of circulating cortisol, Insulin-like growth factor 1 (IGF-1), or testosterone are well documented. To get sufficient numbers of individuals inside the normal and abnormal classes for any of the 3 hormones studied, a biobank containing 655 samples was built from a cohort of 250 sportsmen.

Existing research on automatic prediction within metabolomic data has studied the use of classification algorithms such as Support Vector Machines (SVMs), Random Forests and Partial Least Squares (PLS, \cite{wold66}), and subspace projection methodologies such as Principal Components Analysis (PCA, \cite{hotelling33pca}). In our study we motivate the problem under the framework of bipartite ranking (explained in detail in the next section), and examine the data described above using a selection of state-of-the-art algorithms using a plethora of feature transformations which together account for the functional nature of the data. 

The paper is organised as follows. In Section \ref{sec:background} we detail some of the challenges in making functional prediction and then introduce bipartite ranking as an informative way to view the prediction problem. Section \ref{sec:metabolomicData} details our data corpus and the following section presents an empirical examination of the dataset. We conclude in Section \ref{sec:conclusions}. 

\section{Metabolomics Meets Machine-Learning} \label{sec:background}

It is the purpose of this section to formulate the predictive problems related to metabolomics data from a machine-learning perspective and describe the nature of the input measurements, together with the difficulties raised by their (functional) nature.

\subsection{NMR-based Fingerprints and Predictive Issues}

NMR-based fingerprints are recorded as a continuous spectrum however they are often transformed into a set of features using a bucketing technique. One of the challenges of working with this type of data is that it is high dimensional and the features are correlated and hence learning suffers from the curve of dimensionality (see \cite{abraham2006kernel} for a discussion). The classification based on curves such as those obtained using NMR is called \emph{functional classification} since the observations are sampled functions rather than simple high-dimensional vectors. The analysis of functional data is classed as Functional Data Analysis \cite{ramsay1997functional, ferraty2006nonparametric}. 

One way to deal with the challenges faced with functional data is to use \emph{dimensionality reduction} in which the inputs curves are projected into a finite dimensional subspace. Common examples of dimensionality reduction methods include PCA and PLS. The projection can then be used in conjunction with a classification algorithm to make predictions. Another commonly used approach in signal processing is \emph{filtering}, which thresholds the coefficients of the spectrum that are below a certain base-line. As a simple example, one chooses the first $d$ coefficients with respect to a certain basis. Yet another way to deal with functional data is to use \emph{regularisation}, for example see \cite{hastie1995penalized}. 

\subsection{Wavelets}

As we mentioned, the direct use of statistical methods on raw spectra can result in poor performance, and to address this issue the \emph{wavelet transform} \cite{mallat1999wavelet, daubechies1992ten} can be used. Wavelets map spectra onto a set of frequency components with localisation in space as well as frequency. They can be used to compactly represent a signal, or via thresholding, to denoise signals. An advantage of wavelets over the another similar transform, the Fourier transform \cite{bracewell1986fourier}, is that wavelet coefficients are localised in time and hence frequency changes in the spectrum, such as discontinuities and peaks, can be effectively modelled. In contrast, Fourier analysis is conducted over the whole time period of the spectrum. 

Consider the space of square integrable functions on $[0, 1]$ denoted by $L_2([0,1])$, and denote the observed value of a variable $Z$ at time $t$ using the notation $Z(t)$. The wavelet decomposition finds coefficients  at different times and scales to approximate $L_2$ with respect to a set of orthonormal functions $\psi_{j,k}$, $k=0, \ldots, 2^j-1$, using a \emph{mother wavelet} $\psi_{j,k}(t) = 2^{j/2}\psi(2^j t - k)$ for resolution level or scale $j\geq0$ and translation $k=0, \ldots ,2^j-1$. The space $L_2([0,1])$ can be approximated using a set of closed subspaces $V_0 \subset V_1 \cdots \subset L_2([0,1])$ in which $V_j$ is spanned by $2^j$ orthonormal scaling functions $\phi_{j,k}$, $k=0, \ldots, 2^j-1$. The orthonormal complement between $V_j$ and $V_{j+1}$ is generated by the mother wavelets $\psi_{j,k}, k=0, \ldots, 2^j-1$. It follows that the space $L_2([0,1])$ can be written using the orthonormal basis $\cup_{j \geq 0} \{\psi_{j,k} \}_{k=0,\ldots,2^j-1}$ and $\phi_{0,0}$. Therefore an observation in $Z$ can be written as 
\begin{displaymath} 
Z(t) = \sum_{j=0}^\infty \sum_{k=0}^{2^j-1} \alpha_{j,k} \psi_{j,k}(t) + \mu \phi_{0,0}(t), 
\end{displaymath} 
for $t \in [0,1]$. The coefficients are found using $\alpha_{j,k} = \int_0^1 Z(t) \psi_{j,k}(t) dt \quad \mbox{and} \quad  \mu = \int_0^1 Z(t) \phi_{0,0}(t) dt$, 
which amounts to the projection of a spectra onto the corresponding basis function.  In practice one limits the resolution level with an upper bound $J$ and the approximation of the spectrum is given by 
\begin{displaymath} 
Z(t) \approx \sum_{j=0}^{J-1} \sum_{k=0}^{2^j-1} \alpha_{j,k} \psi_{j,k}(t) + \mu \phi_{0,0}(t). 
\end{displaymath} 
Thus, instead of using the raw spectra features, one can use wavelet coefficients $\alpha_{j,k}$, $j=0, \ldots,J-1$, $k=0, \ldots, 2^j-1$, and $\mu$. Common mother wavelets include Haar, Daubechies and Symmlet wavelets. The latter two have different forms according to the number of \emph{vanishing moments} in which a wavelet with $M$ vanishing moments has $\int t^k \psi(t) dt = 0$ for $k \in \{0, \ldots, M-1 \}$. Wavelets have been used to preprocess metabolomic data in \cite{davis2007adaptive, xia2007integration}. 

\subsection{Prediction Approaches} 

One way of tackling the classification problem using wavelets is to threshold the wavelet coefficients that have the smallest values, which are deemed irrelevant. In hard-thresholding one removes the smallest coefficients and leaves the others unchanged. Soft thresholding in addition reduces the values of the remaining coefficients by the threshold value. A more pragmatic approach is to find those coefficients which are most useful for prediction, for example in \cite{alexandrov2009biomarker}. This effectively corresponds to \emph{feature selection} \cite{guyon06featureExtraction}, which is the problem of selecting a set of features which are relevant to the label being predicted. 

Feature selection is a broad field ranging from individual feature scoring, to subset selection through search, to projection methods into subspaces. We consider combined feature selection and prediction approaches, and a useful technique is to find the minimum distance projection onto a subspace with a penalty on the L0 or L1 norm. Doing so results in a sparse set of projection vectors and hence not all features are used in the prediction of new test examples. In this vein, there have been a plethora of dimensionality reduction methods, for example Sparse PCA \cite{zou2006sparse, aspremont04sparsePCATech, moghaddam2005spectral} and sparse PLS \cite{hoegaerts2004pss, dhanjal2009efficient, momma03sparsepls}. 

The Least Absolute Shrinkage and Selection Operator (LASSO \cite{Tib:96}) is one approach for prediction that has gained popularity in recent years. When working with classification problems one can simply use the predicted real output and then threshold in order to provide a positive or negative label. LASSO is applied to metabolic data in \cite{rohartphenotypic} and compared to bootstrapped LASSO. Other prediction studies of metabolomic data include  \cite{domange08orthologous} which applies PLS Discriminant Analysis (PLS-DA) to MRI associated with orphan neurological disease. A study of of NMR spectra of urine of Streptococcus Pneumoniae in \cite{mahadevan08analysis} demonstrates an improvement of SVMs over PLS-DA.  In \cite{bryan2007using} Random Forests are applied to the metabolic classification problem, with a particular relevance to the important features used in prediction. In contrast to these studies, we examine metabolomic data in the framework of bipartite ranking which has the advantages outlined below. 

\subsection{Bipartite Ranking} 

The metabolomic data studied in this article is labelled with unbalanced classes, for example few positive and many negative labels. In this case it is advantageous to study the bipartite ranking problem as opposed to binary classification. In bipartite ranking one assigns scores to a set of examples labelled in a binary fashion. This is useful, as will later become clearer, since it monitors two error types: the rate of missed detections and false positives. The competitiveness of ranking compared to regression methods such as LogitBoost \cite{friedman2000special} and Kernel Logistic Regression \cite{zhu2002kernel} is demonstrated in \cite{clemencon2011medical}. 

To formalise bipartite ranking consider a set of examples-label pairs denoted by $S = \{(\xv_1, y_1), \ldots, (\xv_n, y_n)\}$ in which examples are denoted $\xv_i \in \mathcal{X}$ and labels using $y_i \in \{-1, +1\}$, $i = 1, \ldots, n$. There exists a scoring function on the set of examples $s : \mathcal{X} \rightarrow \mathbb{R}$  such that $\forall (\xv, \xv') \in \mathcal{X}^2, \xv \leq_s \xv' \Leftrightarrow s(\xv) \leq s(\xv')$. The Receiver Operator Characteristic (ROC) curve $\mbox{ROC}(s, t)$ is defined as $t \in \mathbb{R} \mapsto (P\{s(X) > t \; |\; Y = -1 \}, P\{s(X) > t \; |\; Y = +1 \})$ where $t$ is a threshold on the scoring function. In words, the ROC curve is a plot of the true positive rate (probability of a positive prediction given a positive label) versus the false positive rate (probability of a positive prediction given a negative label) as the threshold value $t$ changes. The ROC curve is informative because it represents the classification errors for all scoring thresholds $t$. Hence, one can for example, choose an acceptable false positive rate and trade off the true positive rate accordingly. 

One way to measure the quality of a ROC curve in a scalar manner, which is useful for optimisation, is using the Area Under the ROC Curve (AUC), defined by $\mbox{AUC}(s) = \int_0^1 \mbox{ROC}(s, \alpha) d\alpha$. A larger AUC corresponds to a better scoring function. It can also be written in terms of the rate of concording pairs, 
\begin{eqnarray*} 
\mbox{AUC}(s) &=& P(s(X) > s(X') | Y=+1, Y'=-1)  \\ 
&+& \frac{1}{2} P(s(X) = s(X') | Y=+1, Y'=-1), 
\end{eqnarray*}
where $(X, Y)$ and $(X', Y')$ are i.i.d. observations.  In bipartite ranking the aim is to find a scoring function with a ROC curve as close to the optimal curve as possible and this implies a small residual AUC with respect to the optimal. In the following subsections we outline the bipartite ranking algorithms: Ranking Forests \cite{clemenccon2013ranking}, Rank SVM \cite{joachims2002optimizing} and RankBoost \cite{freund2003efficient}. 

\subsubsection{Ranking Forests}

\emph{TreeRank Forest} or \emph{Ranking Forests} is a ranking algorithm based on a collection of individual \emph{TreeRank} learners. One first constructs a set of \emph{ranking trees} each denoted $\mathcal{T}_D$ where $D \geq 0$ is the depth of the tree. The root of the tree is labelled $C_{0,0}$ and the non-leaf nodes are given by $C_{d, k}$ for depth $d$ and $0 \leq k \leq 2^{d}$, which has two children $C_{d+1, 2k}$ (left) and $C_{d+1, 2k+1} = C_{d,k} \backslash C_{d+1,2k}$ (right). Given a particular ranking tree the score of an example is computed as 
\begin{displaymath} 
 s_\mathcal{T}(\xv) = \sum_{C_{d,k}: \mbox{terminal cell}} 2^D(1-k/2^d) \cdot \mathbb{I}\{\xv \in C_{d,k}\}.
\end{displaymath}
In other words, the scores ascend going from left to right along the leaf nodes. 

To construct the tree, TreeRank follows a similar idea to that of Classification and Regression Trees (CART, \cite{breiman1984classification}) and splits the data in order to maximise the AUC at each node. At cell $C$ one split the data into two mutually exclusive subsets $C_+$ and $C_- = C \backslash C_+$ so that the error given by: 
\begin{eqnarray*} 
L_{C, \omega}(\Gamma) &=& \frac{2(1-\omega)}{n} \sum_{i=1}^n \mathbb{I}\{\xv_i \in C \backslash \Gamma \} \cdot \mathbb{I}\{y_i = +1\}  \\
& & + \frac{2\omega}{n} \sum_{i=1}^n \mathbb{I}\{\xv_i \in \Gamma \cap C\} \cdot \mathbb{I}\{y_i = -1\}, 
\end{eqnarray*}
(or a penalised/convexified variant) is minimised for a given weight $0 \leq \omega \leq 1 $. One starts with $C_{0,0} = \mathcal{X}$ and the iterates over $d=0, \ldots,D-1$ and $k=0,\ldots,2^d-1$ in the following way: the rate of positives in the current cell $C_{d,k}$ is given by $\alpha(C_{d,k}) = \frac{1}{n_{d,k}} \sum_{i=1}^n \mathbb{I}\{\xv_i \in C_{d,k} \} \cdot \mathbb{I}\{y_i = +1\}$, where $n_{d,k}$ is the number of elements in the cell. One then minimises $L_{C_{d,k}, \omega}(\Gamma)$ with $\omega = \alpha(C_{d,k})$. This provides the split into the two child cells as $C_{d+1,2k} = C_+$ and $C_{d+1, 2k+1} = C_{d,k} \backslash C_{d+1, 2k}$. The splitting rule can be formed using any cost-sensitive classifier (see \cite{hastie2005elements}) and termed a LeafRank algorithm. The algorithm is shown to be a piecewise linear interpolation scheme of the optimal ROC curve. 

To combine ranking trees one uses bootstrap aggregating (bagging) in conjunction with feature randomisation. In essence, one draws a number $B$ of bootstrap samples (random sampling with replacement) and then selects a random subset of the features of the examples for each sample and these are used to train $B$ ranking trees. There are a number of methods to aggregate the scoring functions of the set of trees, for example by computing the median ranking with respect to a distance metric between rankings. One can also simply compute the mean scoring function. 

\subsubsection{RankBoost} 

RankBoost is based on  AdaBoost \cite{FreundSchapire:95}, and works by combining many ``weak'' rankings into a single accurate one. At iteration $t$ one maintains a distribution matrix $\Dm_t \in \mathcal{X}\times\mathcal{X}$ which emphasises the importance of orderings on particular pairs of examples. We replicate the pseudo-code for RankBoost in Algorithm \ref{alg:rankboost} for clarity.  

\begin{algorithm}
\caption{RankBoost pseudo-code}
\label{alg:rankboost}
\begin{algorithmic}[1]
\STATE Input: Initial distribution matrix $\Dm$
\STATE $\Dm_0 = \Dm$ 
\FOR{$t=1,\ldots,T$} \label{lin:for}
\STATE Train weak learner using $\Dm_t$ and choose $\alpha_t \in \mathbb{R}$
\STATE Get weak ranking $h_t: \mathcal{X} \rightarrow \mathbb{R}$ 
\STATE Set $\Dm_{t+1}(\xv_0, \xv_1) = \frac{\Dm_{t}(\xv_0, \xv_1) \exp(\alpha_t(h_t(\xv_0) - h_t(\xv_1)))}{\Zm_t}$ where $\Zm_t$ is a normalisation constant
\ENDFOR
\STATE Output: $H(\xv) = \sum_{i=1}^T \alpha_i h_i(\xv)$
\end{algorithmic}
\end{algorithm}

The key idea is that the distribution matrix $\Dm_t$ is modified at each iteration so as to stress the importance of certain pair orderings. If $\xv_1$ is to be ranked higher than $\xv_0$ and $\alpha_t > 0$ then if $h_t(\xv_1) > h_t(\xv_0)$ the update rule on $\Dm_{t+1}(\xv_0, \xv_1)$ will increase its value. Further details of how to set $\alpha_t$ and some algorithms for weak learners are given in \cite{freund2003efficient}. 

\subsubsection{Ranking SVM} 

Finally we overview the Ranking SVM which is a way to maximise a functional closely related to an similarity measure between two rankings known as the \emph{Kendall tau}. Let $P$ be the number of concording pairs and $Q$ be the number of discording pairs between two rankings $h$ and $h'$, then Kendall tau is defined as $\tau(h, h') = (P-Q)/(P+Q)$. The Ranking SVM approach is based on the idea of maximising a margin as in SVMs whilst minimising Kendall tau. In particular the following convex optimisation is used, 
\begin{displaymath} 
\begin{array}{l l}
 \min &  \frac{1}{2}\|\wv\|^2 + C \sum \varepsilon_{i, j, k} \\ 
\st & \forall (\xv_i, \xv_j) \in h_1 : \langle \wv, \xv_i \rangle \geq  \langle \wv, \xv_j \rangle + 1 - \varepsilon_{i, j, 1} \\ 
& \cdots \\ 
& \forall (\xv_i, \xv_j) \in h_n : \langle \wv, \xv_i \rangle \geq  \langle \wv, \xv_j \rangle + 1 - \varepsilon_{i, j, n} \\ 
& \forall i, j, k : \varepsilon_{i, j, k} \geq 0,  
\end{array}
\end{displaymath}
in which $C$ is a user-defined variable which trades off the margin size against training error,  $\varepsilon_{i, j, k}$ are slack variables and $\wv$ is a weight vector. Here, $h_1, \ldots, h_n$ are rankings such that if $(\xv_i, \xv_j) \in h_i$ then $\xv_i$ is ranked higher than $\xv_j$. As with SVMs one can use a weight vector which a linear combination of the examples and this allows for the use of kernels for non-linear feature mappings. 

\section{Metabolomic Data} \label{sec:metabolomicData}

This exploratory retrospective study was performed on sera obtained from high level or professional cyclists. Samples were stored in a biobank built in the frame of a large longitudinal medical follow-up of cyclists. This biobank was put under the responsibility of the French Federation of Cycling (FFC). The 3 hormones cortisol, IGF-1, and testosterone, were routinely assayed to detect endocrine disruptions as described for cortisol \cite{guinot2007value}. From the analysis of variations of cortisol, testosterone and IGF-1 concentrations, we have focussed on 3 types of substances suspected to induce endocrine disruptions, which have indirect repercussions on the general metabolism as it is accessible at the serum level by a NMR-based metabolomic observation. More precisely, use of prohibited compounds can be suspected from variation of cortisol, testosterone or IGF-1 levels in serum from what is considered as normal values. 

Most of samples were collected in 2001 and 2002 sporting seasons as previously described by Guinot et al. \cite{guinot2007value}. 655 samples were chosen to constitute this working biobank, which was built from 250 sportsmen submitted to a longitudinal medical follow-up. All samples were obtained anonymously to comply with the French legislation concerning the scientific follow-up of human populations as published previously.

\subsection{Feature Generation and Processing} 

For the complete set of 655 NMR spectra the range of concentrations of IGF-1, cortisol and testosterone are shown in Table \ref{tab:concentrations}. Each spectrum is composed of 950 measurements of resonance at differing frequencies. Most of the samples for testosterone and IGF-1 fall within the normal range. With cortisol we see close numbers of normal and high sample concentrations. Rather that working solely on the raw spectra, we use PCA with 100 eigenvectors to decorrelate features, and also Haar, Daubechies-4 (Db4) and Daubechies-8 (Db8) multilevel wavelets with $J=10$ levels, resulting in 100, 953, 1012, 1093 features respectively. In order to use a discrete wavelet transform the input feature vector much be of length $2^p$ where $p$ is an integer, and we pad our features with the border values for them to satisfy this requirement.  

\begin{table}[ht]
\begin{center} 
\begin{tabular}{l | l l | l l | l l}
\hline			
& \multicolumn{2}{c}{low} & \multicolumn{2}{c}{normal} & \multicolumn{2}{c}{high} \\
& Conc. & Freq. & Conc. & Freq. & Conc. & Freq.  \\
\hline 
testosterone & 0-3  & 78  & 3-9  & 500  & 9-13  & 57  \\
cortisol & 0-89 & 62 & 89-255 & 303 & 255-573 & 268  \\
IGF-1 &  0-200 & 71 & 200-441 & 472 & 441-781 & 88 \\
\hline  
\end{tabular}
\end{center}
\label{tab:concentrations}
\caption{The concentration ranges (ng/ml) for each hormone into the categories: low, normal and high. Also shown is the number of individuals in the sample within the corresponding range.} 
\end{table} 

Since hormone concentrations are known to vary with age, we include the age in years as an additional feature in all feature sets. Any missing values in the label values result in the corresponding example being omitted, however only a maximum of 24 examples contained missing values depending on the label under consideration. There were several age values that were missing and we replaced these with the mean age from the non-missing values. 

\section{Computational Analysis} \label{sec:compAnalysis} 

\begin{figure}[ht]
\begin{center} 
\includegraphics[width=0.45 \linewidth]{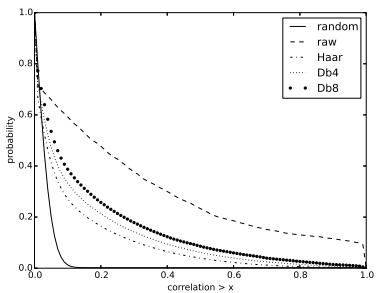}
\end{center}
\caption{The correlations of the spectra and wavelet coefficients.} 
\label{fig:correlations}
\end{figure}

Before ranking the spectra, we study the correlation present in the current set of features to confirm the need for feature selection. To do so we compute a correlation matrix $\Rm$ for each feature set, where $\Rm_{ij} = \sum_k \Xm_{ki}\Xm_{kj} / \sqrt{\sum_k \Xm_{ki}^2 \sum_k \Xm_{kj}^2}$ for zero-mean examples. Figure \ref{fig:correlations} shows the number of pairs in the correlation matrices for each feature set above different thresholds.  The key observation is that the raw spectrum contains a large number of features which are closely correlated. There is an initial drop in the beginning of the curve since the spectra are zeroed out for the range corresponding to water. Nearly 10\% of the pairs of features have a correlation greater or equal to 0.99. Wavelet features are less correlated than the raw spectrum ones, however both are still significantly more correlated than features which are uniformly randomly generated.  

\begin{figure}[ht]
\begin{center} 
\subfigure[Filter error]{\includegraphics[width=0.45 \linewidth]{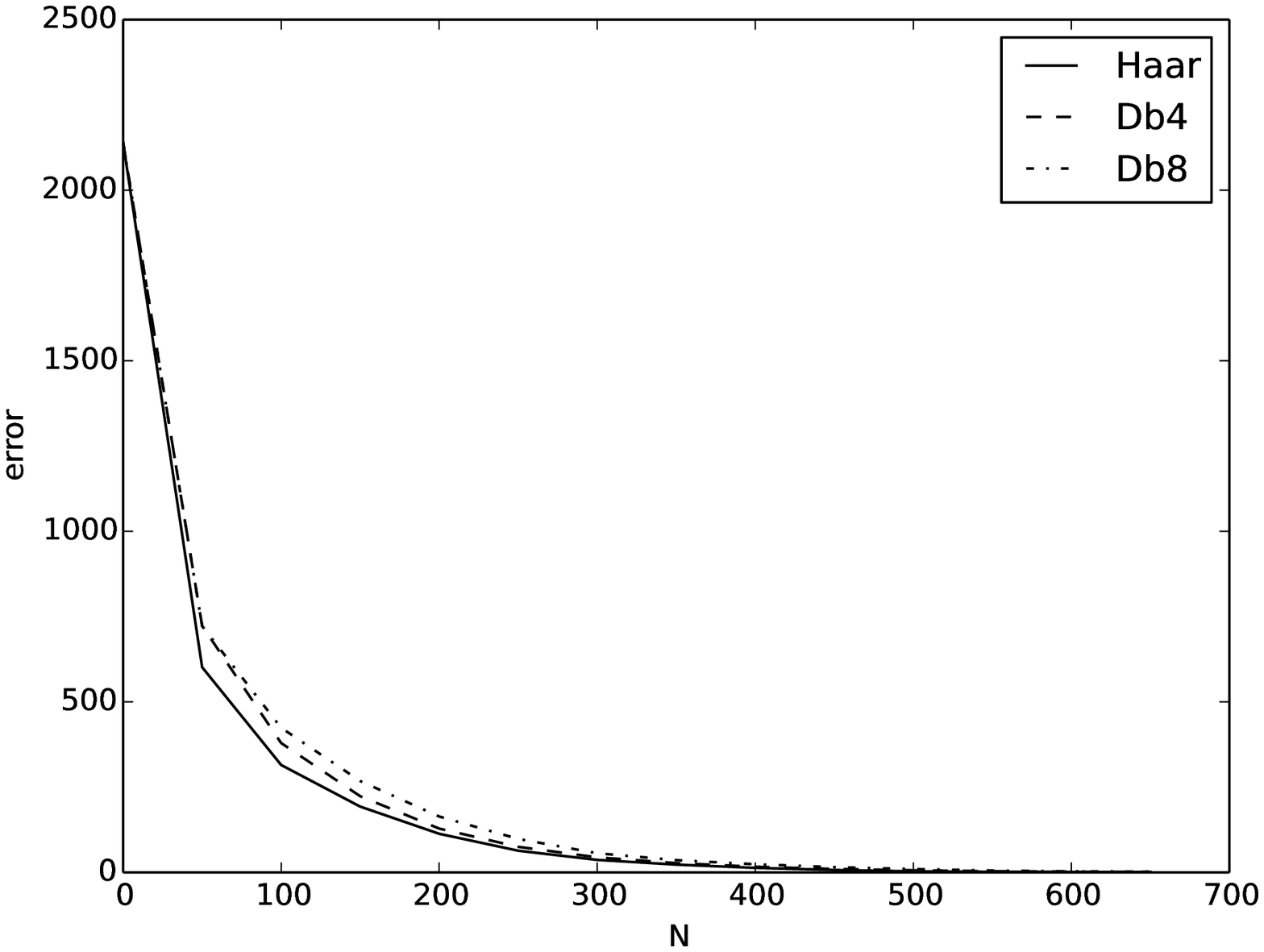}\label{subfig:filterErrors}} 
\subfigure[Reconstructed spectrum]{\includegraphics[width=0.45 \linewidth]{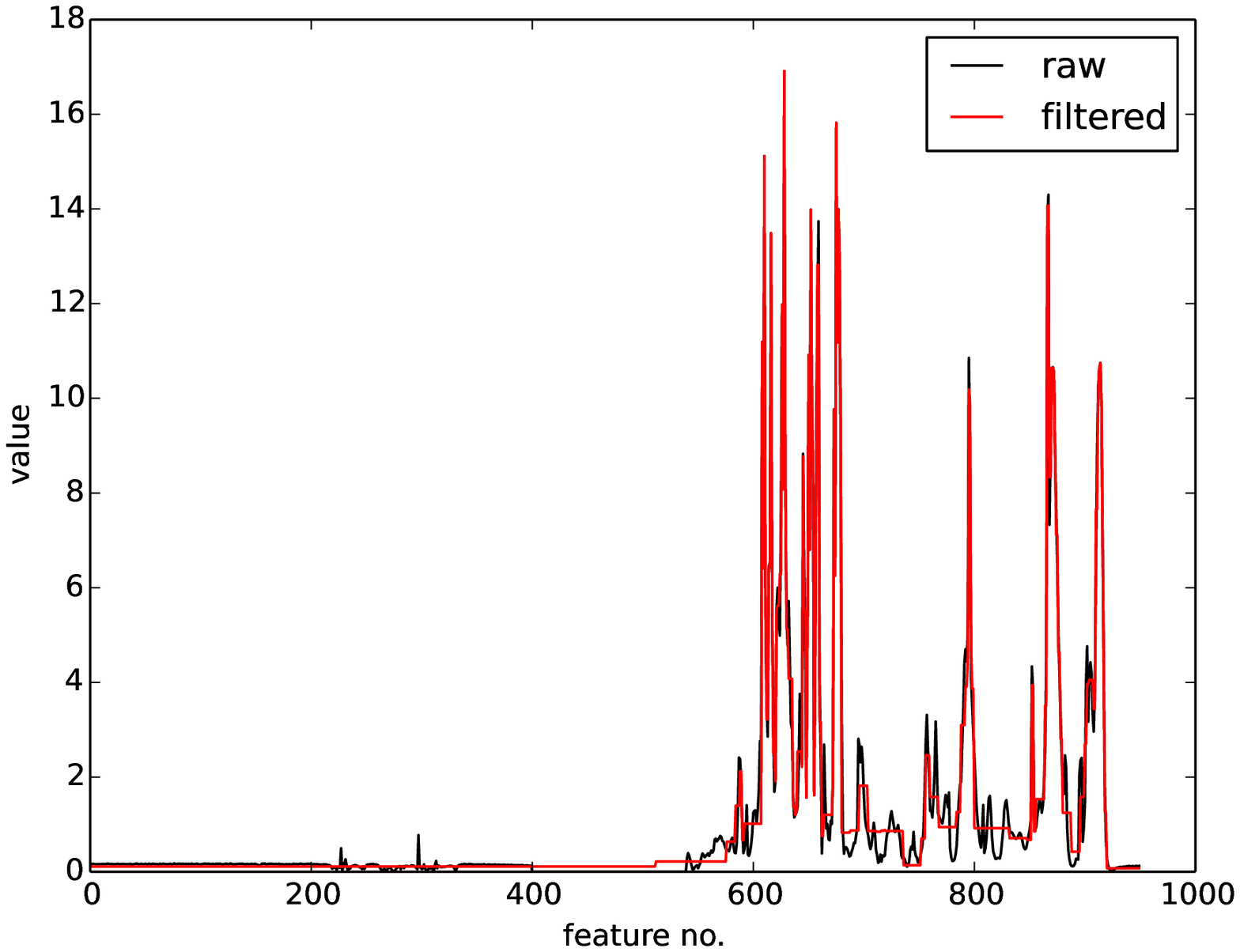}\label{subfig:reconstructErrors}} 
\end{center}
\caption{An illustration of the errors introduced using wavelet filtering.} 
\label{fig:filterErrors}
\end{figure}

Next, PCA is applied to decorrelate the features. The largest 100 eigenvectors of the covariance matrix are computed for the raw and wavelet features, with the covariance matrix defined using $\Cm = \frac{1}{n}\Xm^T\Xm$ for centred examples. We observed that the first 100 eigenvalues model 99.1\% of the total variance for the raw spectra and approximately 99.99\% of the variance of the wavelets. This implies that the features live predominantly in a low dimensional subspace. Adjacent to this analysis is the reconstruction of the spectra using a filtered subset of size $N$ of the wavelet coefficients. In Figure \ref{subfig:filterErrors} we observe how the error in the filtered spectra vary with $N$ by measuring the Frobenius norm difference between the real and approximated spectra, where the Frobenius norm of $\Am$ is $\|\Am\|_F^2 = \sum_{ij}\Am_{ij}^2$. The initial error is 2139.4 (the norm of the original spectra) and this falls to 372 on average for $N=100$. Haar wavelets result in the lowest error and the highest are found with the Db8 wavelets. In Figure \ref{subfig:reconstructErrors} we show an example of the reconstruction of a spectrum after filtering Haar wavelet coefficients. One can see that minor details are removed and large peaks are preserved which is one would expect when performing filtering. We later use an L1 regularised SVM in conjunction with TreeRank Forests for feature selection based on predictive ability, rather than variance as with filtering for example. 

\subsection{Learning Setup} 

As previously discussed, a useful way of viewing the problems above is as a set of bipartite ranking ones over the class of the corresponding hormone (low, normal and high concentrations). The aim is to see  which feature sets are more predictive and also to examine the individual features which are useful for learning.  We use the following sets of features: raw spectra (denoted \texttt{raw}), the PCA transformed data (\texttt{pca}), the Daubechies wavelet of length 4 and 8 (denoted by \texttt{Db4}, \texttt{Db8} respectively), and the Haar wavelet (\texttt{Haar}). Each set of features is standardised by centring and scaling each feature to have unit norm. For ranking the following algorithms are used: Ranking Forests (\texttt{TRF}), RankBoost (\texttt{RB}) and the Ranking SVM (\texttt{RSVM}). TreeRank is excluded from the experimentation since in both our preliminary work and in \cite{clemenccon2013ranking} Ranking Forests consistently improves over TreeRank. 

\subsection{Bipartite Ranking} 

Each approach is evaluated by measuring AUC using 3-fold cross validation and averaging the results. Within each of these folds, model selection is performed using an inner 5-fold cross validation loop.  At each cross validation split we stratify the labels to ensure that each split has a distribution of labels similar to the others. 

The parameters for the ranking methods are chosen by selecting those with the highest AUC in conjunction with the inner cross validation loop. For RankBoost we choose the number of iterations from $\{10, 50, 100\}$ and the number of weak rankers from $\{5, 10, 20\}$. With the Ranking SVM we use the Radial Basis Function (RBF) kernel defined by $\kappa(\xv, \xv') = \exp(-\gamma \|\xv - \xv'\|^2)$. Parameters are chosen as $C \in \{2^{0}, 2^{1}, 2^2\}$ and $\gamma \in \{2^{-3}, 2^{-2}, 2^{-1}\}$. For Ranking Forests the minimum split size of a node is set to 50, the maximum depth is fixed at 10 and we use 20 trees. Randomisation improves the performance of TreeRank Forest and so we use bootstrap sampling of the training set on sizes $\{0.5, 0.75, 1.0\}$ of the number of training examples, and similarly we select random feature proportions $\{0.5, 0.75, 1.0\}$. In conjunction with TreeRank Forest we use three LeafRank algorithms: CART, RBF SVM  and the L1 penalised SVM denoted \texttt{CART}, \texttt{RBF} and \texttt{L1} in the results. The latter is particularly interesting in terms of finding the relevant features. For each LeafRank algorithm we perform model selection using 3-fold cross validation. For CART we select the maximum depth from $\{2, 4, 8\}$ and fix the minimum split size to 30. In the case of the SVM, $C \in \{2^{-5}, 2^{-3}, \ldots, 2^7 \}$ and with the RBF SVM, $\sigma \in \{2^{-5}, 2^{-3}, \ldots, 2\}$.
\begin{table*}[ht]
\small
\begin{center} 
\begin{tabular}{l | l l l | l l l | l l l }
\hline	
& \multicolumn{3}{|c|}{IGF-1} & \multicolumn{3}{|c|}{cortisol}  &  \multicolumn{3}{|c}{testosterone}\\ 
& low & normal & high & low & normal & high & low & normal & high \\
\hline	
& \multicolumn{8}{c}{raw} \\ 
\hline  
CART-TRF             & .79 (.01) & .72 (.06) & .71 (.01) & .63 (.07) & .68 (.03) & .67 (.04) & .66 (.04) & .57 (.01) & .63 (.03)\\
L1-TRF               & .79 (.02) & .69 (.03) & \textbf{.79} (.05) & .71 (.13) & .73 (.01) & \textbf{.78} (.03) & \textbf{.69} (.07) & \textbf{.60} (.03) & .67 (.06)\\
RBF-TRF              & \textbf{.80} (.02) & .70 (.01) & .73 (.03) & .71 (.10) & \textbf{.74} (.02) & .76 (.04) & .68 (.06) & .57 (.02) & .68 (.04)\\
RB                   & .74 (.04) & \textbf{.72} (.03) & .77 (.01) & .71 (.08) & .72 (.01) & .76 (.02) & .61 (.07) & .58 (.04) & \textbf{.71} (.03)\\
RSVM                 & .71 (.03) & .63 (.02) & .71 (.05) & \textbf{.74} (.10) & .69 (.01) & .73  (.05) & .66 (.04) & .56 (.02) & .70 (.01)\\
\hline  
& \multicolumn{8}{c}{pca} \\ 
\hline  
CART-TRF             & .68 (.01) & .67 (.04) & .68 (.02) & .65 (.02) & .62 (.02) & .61 (.01) & .62 (.06) & .50 (.04) & .48 (.02)\\
L1-TRF               & \textbf{.76} (.01) & .68 (.01) & \textbf{.79} (.04) & .69 (.11) & .69 (.03) & .75 (.04) & .63 (.06) & .57 (.02) & .67 (.04)\\
RBF-TRF              & .74 (.03) & .65 (.02) & .78 (.02) & \textbf{.70} (.08) & \textbf{.72} (.01) & \textbf{.77} (.04) & \textbf{.66} (.06) & \textbf{.61} (.02) & \textbf{.70} (.02)\\
RB                   & .71 (.02) & \textbf{.68} (.05) & .73 (.02) & .54 (.04) & .68 (.03) & .72 (.02) & .60 (.09) & .53 (.03) & .60 (.03)\\
RSVM                 & .69 (.01) & .62 (.02) & .74 (.06) & .68 (.08) & .70 (.02) & .76 (.05) & .64 (.05) & .59 (.01) & .67 (.02)\\
\hline  
\end{tabular} 
\end{center} 
\caption{The AUC values of the ranking methods using the raw and PCA spectra with standard deviations in parentheses and best results in bold.} 
\label{tab:rawFeatures}
\end{table*} 

\begin{table*}[ht]
\small
\begin{center} 
\begin{tabular}{l | l l l | l l l | l l l }
\hline	
& \multicolumn{3}{|c|}{IGF-1} & \multicolumn{3}{|c|}{cortisol}  &  \multicolumn{3}{|c}{testosterone}\\ 
& low & normal & high & low & normal & high & low & normal & high \\
\hline
& \multicolumn{8}{c}{Db4} \\ 
\hline
CART-TRF             & .76 (.03) & .70 (.02) & .71 (.02) & .67 (.09) & .64 (.04) & .70 (.04) & .61 (.04) & .58 (.02) & .61 (.03)\\
L1-TRF               & .74 (.03) & .70 (.01) & .74 (.04) & .69 (.09) & .68 (.02) & .74 (.03) & .66 (.04) & .57 (.01) & .70 (.08)\\
RBF-TRF              & .77 (.03) & \textbf{.71} (.01) & \textbf{.75} (.06) & .66 (.06) & \textbf{.71} (.01) & .76 (.03) & .66 (.02) & \textbf{.59} (.02) & .68 (.04)\\
RB                   & .73 (.02) & .71 (.01) & .75 (.04) & .65 (.04) & .70 (.01) & .74 (.01) & .63 (.02) & .55 (.03) & .66 (.06)\\
RSVM                 & \textbf{.78} (.03) & .69 (.01) & .73 (.05) & \textbf{.70} (.09) & .70 (.01) & \textbf{.77} (.04) & \textbf{.69} (.03) & .57 (.02) & \textbf{.72} (.04)\\
\hline
& \multicolumn{8}{c}{Db8} \\ 
\hline	
CART-TRF             & .76 (.02) & \textbf{.72} (.02) & .70 (.03) & .64 (.08) & .65 (.01) & .68 (.02) & .61 (.05) & .56 (.05) & .57 (.04)\\
L1-TRF               & .75 (.03) & .72 (.01) & \textbf{.76} (.03) & .66 (.08) & .68 (.01) & .74 (.01) & .64 (.05) & \textbf{.58} (.03) & .65 (.03)\\
RBF-TRF              & .77 (.03) & .71 (.01) & .74 (.05) & .69 (.10) & .70 (.01) & .76 (.03) & .66 (.05) & .57 (.01) & .67 (.02)\\
RB                   & .73 (.03) & .67 (.02) & .73 (.06) & .62 (.08) & .68 (.04) & .69 (.02) & \textbf{.68} (.02) & .58 (.02) & .64 (.05)\\
RSVM                 & \textbf{.78} (.02) & .69 (.01) & .72 (.05) & \textbf{.70} (.10) & \textbf{.70} (.01) & \textbf{.76} (.04) & .68 (.02) & .56 (.02) & \textbf{.69} (.03)\\
\hline  
& \multicolumn{8}{c}{Haar} \\ 
\hline 
CART-TRF             & .72 (.05) & \textbf{.72} (.01) & .68 (.07) & .62 (.03) & .65 (.02) & .71 (.00) & .61 (.03) & .52 (.03) & .58 (.08)\\
L1-TRF               & .76 (.05) & .70 (.01) & .73 (.06) & .68 (.06) & .69 (.01) & .74 (.03) & .64 (.03) & .58 (.01) & .64 (.09)\\
RBF-TRF              & .77 (.03) & .71 (.01) & \textbf{.75} (.06) & .70 (.08) & .70 (.01) & \textbf{.77} (.02) & .64 (.01) & \textbf{.61} (.02) & .66 (.03)\\
RB                   & .77 (.05) & .70 (.03) & .74 (.05) & .60 (.07) & .70 (.04) & .72 (.01) & .60 (.03) & .52 (.02) & .66 (.03)\\
RSVM                 & \textbf{.78} (.03) & .69 (.01) & .74 (.05) & \textbf{.71} (.10) & \textbf{.70} (.01) & .76 (.04) & \textbf{.67} (.03) & .57 (.01) & \textbf{.72} (.04)\\
\hline  
\end{tabular} 
\end{center} 
\caption{The AUC values of the ranking methods using the wavelet features with standard deviations in parentheses and best results in bold.} 
\label{tab:waveletFeatures}
\end{table*} 

Results are presented in Tables \ref{tab:rawFeatures} and \ref{tab:waveletFeatures}. Consider first the best results for each concentration, and surprisingly these are given by the raw spectra. Evidently, the high correlations present in these features do not adversely affect the results. In contrast, moving to PCA gives slightly worse performance in general. We see that high and low concentrations of the hormones are easier to rank than those in the normal range as we might expect since the predicted scoring functions for the concentrations are bisected more naturally in these cases. IGF-1 is the easiest hormone to rank followed by cortisol and then testosterone. When we look at the ranking methods for the raw spectra, \texttt{L1-TRF} gives AUCs closest to the best in every case, perhaps due to its ability to perform feature selection in addition to ranking. Notice that the ranking SVM performs well on the wavelet features, improving over the other methods in most cases. 
 
 \addtolength{\tabcolsep}{+1mm}
 
\begin{table*}[ht]
\begin{center} 
\begin{tabular}{l l | l l | l l }
\multicolumn{2}{c |}{IGF-1} & \multicolumn{2}{|c|}{cortisol}  &  \multicolumn{2}{|c}{testosterone}\\ 
low & high & low  & high & low & high \\
\hline
950 & 950 & 567 & 850 & 235 & 711\\
728 & 777 & 694 & 801 & 321 & 677\\
919 & 722 & 547 & 892 & 887 & 728\\
650 & 808 & 315 & 587 & 645 & 920\\
154 & 554 & 606 & 263 & 566 & 695\\
734 & 305 & 305 & 713 & 732 & 815\\
676 & 247 & 950 & 883 & 304 & 785\\
228 & 698 & 548 & 746 & 757 & 696\\
719 & 645 & 884 & 299 & 746 & 923\\
714 & 732 & 709 & 268 & 772 & 620\\
\hline
\end{tabular} 
\end{center} 
\caption{The 10 most relevant features used by \texttt{L1-TRF} on the raw spectra.} 
\label{tab:mostRelevantFeatures}
\end{table*} 

Next we attempt to learn about the most important features used for TreeRank Forest based on an approach given in \cite{clemenccon2011adaptive}. We focus on \texttt{L1-TRF} since it gives strong results using the raw spectra relative to the other ranking methods. To find important features we compute for each ranking tree a marginal gain in AUC given by splitting at an internal node, denoted by $\Delta \mbox{AUC}(m)$ for a node $m$. One uses the square of the marginal AUC to weight the weight vector of the SVM and then sums these quantities to give an indication of the overall importance of each feature for each tree. Formally, the vector of feature importances $\hat{\wv}$ can be written as 
\begin{displaymath} 
\hat{\wv} = \sum_{\mbox{internal nodes $m$}} \left(\Delta \mbox{AUC}(m)\right)^2 \cdot |\wv_m|,  
\end{displaymath}
where $\wv_m$ is the weight vector for node $m$ and $|\cdot|$ is the element-wise absolute value of a vector. For a forest of trees, one simply sums the weight vector for each tree and then normalises so that the resulting vector has unit norm. The resulting zero-based indices are presented in Table \ref{tab:mostRelevantFeatures} for high and low hormone concentrations, since these resulted in the highest AUC scores in Table \ref{tab:waveletFeatures}. Of note here is that there is little overlap in terms of the features used for the same hormone aside from the feature corresponding to age, 950, in the case of IGF-1. Age also appears to useful for ranking concentrations of cortisol. When we look at the first 100 features then there are 35, 31 and 24 in common for low and high concentrations of IGF-1, cortisol and testosterone respectively. This appears to point to the relevance of using difference ranking schemes to detect abnormal concentrations. We leave a more medically centred interpretation for informed readers. 

\section{Conclusions} \label{sec:conclusions}

We explored the scoring problem in metabolomics in the context of predicting elevated metabolite presence in blood serum analysed using NMR spectra. Such a problem was more informatively split into multiple bipartite ranking problems as compared to the traditional classification or regression approach. On a cohort of 655 NMR spectra of French professional or near professional cyclists, the concentration ranges of testosterone, cortisol and IGF-1 were predicted using bipartite ranking methods in an extensive computational analysis. We showed that the raw spectra were more predictive in conjunction with the ranking methods relative to those generated using PCA and wavelet features. Finally we used a novel method to study feature important and highlighted the most important features used for ranking with the L1-penalised SVM TreeRank Forest. 

\section*{Acknowledgements} 
The authors are grateful to Alain Paris of AgroParisTech for granting access to the metabolomics database.

\bibliographystyle{abbrv}
\bibliography{references,my-references}

\end{document}